\documentclass[twocolumn,preprintnumbers,amsmath,amssymb]{revtex4}
\usepackage{dcolumn}
\usepackage{bm}
\usepackage{graphicx,subfigure,xcolor}
\usepackage{braket}

\usepackage{comment}
\usepackage{xcolor}

\newcommand{\wo}{\omega_0}
\newcommand{\wk}{\omega_k}
\newcommand{\wj}{\omega_j}

\newcommand{\ewo}{e^{i\omega_0 t}}
\newcommand{\ewom}{e^{-i\omega_0 t}}

\newcommand{\ewkm}{e^{-i\omega_k t}}

\newcommand{\skj}{\sum_{kj}}
\newcommand{\Ckj}{\mathcal{C}_{kj}}
\newcommand{\Ckju}{\mathcal{C}_{kj}^1}
\newcommand{\Ckjd}{\mathcal{C}_{kj}^2}

\newcommand{\Ckjl}{\mathcal{C}_{kj}^l}

\newcommand{\bd}{b^{\dag}}

\newcommand{\adk}{a^{\dag}_k}
\newcommand{\adj}{a^{\dag}_j}
\newcommand{\aj}{a_j}
\newcommand{\ak}{a_k}

\newcommand{\aku}{a_{k,1}}

\newcommand{\akd}{a_{k,2}}

\newcommand{\adkl}{a^{\dag}_{k,l}}
\newcommand{\adjl}{a^{\dag}_{j,l}}
\newcommand{\ajl}{a_{j,l}}
\newcommand{\akl}{a_{k,l}}

\begin{document}

\title{Nonequilibrium dressing in a cavity with a movable reflecting mirror}

\author{Federico Armata$^1$\footnote{federicoarmata@msn.com}}

\author{M. S. Kim$^1$\footnote{m.kim@imperial.ac.uk}}

\author{Salvatore Butera$^{2}$\footnote{sb469@hw.ac.uk}}

\author{Lucia Rizzuto$^{3,4}$\footnote{lucia.rizzuto@unipa.it}}

\author{Roberto Passante$^{3,4}$\footnote{roberto.passante@unipa.it}}

\affiliation{$^1$QOLS, Blackett Laboratory, Imperial College London, London SW7 2BW, United Kingdom}
\affiliation{$^2$SUPA, Institute of Photonics and Quantum Sciences, Heriot-Watt University, Edinburgh EH14 4AS, United Kingdom}
\affiliation{$^3$Dipartimento di Fisica e Chimica, Universit\`{a} degli Studi di Palermo, Via Archirafi 36, I-90123 Palermo, Italy}
\affiliation{$^4$INFN, Laboratori Nazionali del Sud, I-95123 Catania, Italy}
\affiliation{$^5$Korea Institute of Advanced Study, Dongdaemun-gu, Seoul, 02455, South Korea}

\begin{abstract}
We consider a movable mirror coupled to a one-dimensional massless scalar field in a cavity. Both the field and the mirror's mechanical degrees of freedom are described quantum mechanically, and they can interact with each other via the radiation pressure operator. We investigate the dynamical evolution of mirror and field starting from a nonequilibrium initial state, and their local interaction which brings the system to a stationary configuration for long times. This allows us to study the time-dependent dressing process of the movable mirror interacting with the field, and its dynamics leading to a local equilibrium dressed configuration. Also, in order to explore the effect of the radiation pressure on both sides of the movable mirror, we generalize the effective field-mirror Hamiltonian and previous results to the case of two cavities sharing the same mobile boundary. This leads us to address, in the appropriate limit, the dynamical dressing problem of a single mobile wall, bounded by a harmonic potential, in the vacuum space.
\end{abstract}

\maketitle

\section{\label{sec:Introduction}Introduction}

One of the most fascinating predictions of quantum field theory is the existence of vacuum field fluctuations and related field energy densities. This peculiarity of a quantum field has many observable consequences such as Casimir and Casimir-Polder forces, the Lamb shift and the spontaneous emission of radiation \cite{milonni-book}. Casimir forces are electromagnetic forces of quantum origin, usually attractive, between metallic or dielectric macroscopic objects placed in the vacuum space, even at zero temperature. They originate whenever a field is confined in a specific geometric configuration, which requires imposing boundary conditions on the field, and are ultimately related to a dependence of the vacuum energy from the geometric configuration \cite{casimir-physics}. Remarkable new effects arise when the boundary conditions on a field, or some relevant parameter of the system, change nonadiabatically in time. In this case, the theory predicts emission of pairs of photons from the vacuum \cite{Moore70,Dodonov10}. This effect, known as the dynamical Casimir effect (DCE), has been theoretically investigated in different cases such as nonstationary media in a cavity \cite{Dodonov1993}, perfectly reflecting moving mirrors in a three-dimensional cavity with Dirichlet \cite{Crocce2001} or Neumann-Robin boundary conditions \cite{Crocce2002, Mintz2006}, and partially transmitting mirrors \cite{Haro2006}. Very recently, the dynamical Casimir effect has been observed in different physical systems, for example superconducting circuits \cite{Wilson11} and Bose-Einstein condensates \cite{Jaskula12}. A related dynamical Casimir-Polder effect in a pure quantum electrodynamics framework, exploiting an optomechanical coupling between Rydberg atoms and an oscillating mirror in the near-field regime, has recently been proposed \cite{ABCNPRRS14}. It is now recognized that dynamical or nonequilibrium situations are an excellent platform to explore new physical effects in quantum electrodynamics. In this spirit, dynamical Casimir-Polder interactions have recently been investigated to study the dynamics of an atom-wall system, initially in a nonequilibrium state, before it reaches a stationary configuration \cite{VP08,MVP10,Armata16}. During the transient dynamical dressing, Casimir-Polder forces can be much more intense than in the stationary case and oscillate from attractive to repulsive. Also, subtle conceptual questions concerning with the approach to the equilibrium configuration arise \cite{Armata16,BPRB16}. On the other hand, conceptually related problems such as the response of a quasiparticle to a fast external force have been addressed both theoretically \cite{DuqueGomez-Sipe12} and experimentally \cite{Gaal-Kuhen07}, as well as the observation of the time-dependent dressing of a quasiparticle in condensed matter physics \cite{Hase-Kitajima03}; also, the onset of effective mass in Bose-Einstein condensates has been recently observed \cite{Chang-Potnis14}. All this makes very remarkable investigating the time-dependent dressing of various physical systems.

Another related aspect is when a boundary confining a field is allowed to move and its mechanical degrees of freedom are treated quantum mechanically, and it is thus subjected to quantum fluctuations of its position. In this case there is an effective interaction between the field and the mechanical degrees of the wall (phonons) and an effective interaction between the field modes, mediated by the movable wall \cite{Law1995}.
Recently, it has been investigated how the quantum description of a moving cavity mirror, and related mirror's position fluctuations, can influence the vacuum energy densities of the electromagnetic field inside the cavity and related Casimir energies \cite{Butera2013, Armata2015}. It was found that the effect of the mirror's motion becomes more and more relevant with decreasing mirror's mass. The inclusion of the quantum-mechanical degrees of freedom of a mobile boundary has also been considered with different aims, for example to study the mirror decoherence via the dynamical Casimir effect  \cite{Dalvit2000, Neto2000}, and, more recently, to investigate the role of the vacuum friction and of the anti-correlation properties of the quantum vacuum in confining the mirror's position fluctuations \cite{Wang2014, Wang2015, Volovik2014}. These works can also be of a broader interest, for example in the framework of recent developments in the rapidly growing field of cavity optomechanics, which studies the coupling of optical fields and mechanical degrees of freedom. Indeed, nowadays in optomechanical experiments it is possible to build optomechanical cavities with movable mirror of masses of the order of $10^{-11}$Kg or even less \cite{aspelmeyer2014} and reach temperatures of the order of a few millidegrees Kelvins \cite{Kippenberg2008}. These remarkable experimental achievements let us hope that, in the near future, the exploration of vacuum effects arising from a quantum description of the boundaries acting on fields will be experimentally accessible. Position fluctuations of a boundary are also important in smearing out well-known divergences of the electromagnetic energy density at the boundary \cite{PRS13,Bartolo15}, with possible relevance also for gravitational effects, because the field energy density couples to gravity.

In this paper we consider the interaction of a one-dimensional scalar (or electromagnetic) field with a movable mirror placed inside a perfectly conducting cavity. The mechanical degrees of freedom of the moving boundary are included in the system dynamics according to the Hamiltonian formulation developed in Ref. \cite{Law1995}. In the equilibrium configuration, because of the quantum description of the field-mirror interaction, both field and mirror contain virtual excitations; i.e. the ground state of the interacting system is dressed. This dressed situation has recently been studied in \cite{Butera2013} for a one-dimensional massless scalar field, and successively extended to a one-dimensional electromagnetic field and a three-dimensional massless scalar field in \cite{Armata2015}. Here, by using a framework well-established in the context of the Casimir-Polder forces between atoms \cite{PP03} or between an atom and a conducting mirror \cite{VP08,MVP10,Armata16}, we investigate the dynamical (time-dependent) dressing process bringing the field-mirror system, initially in a dynamical out of equilibrium configuration, to a dressed configuration showing a local equilibrium. This allows us to find how field and mirror exchange virtual excitations before reaching an equilibrium state. By calculating the local dynamical energy shift of the system we are able to explore the transient dynamical dressing process of the wall and the features of approaching the equilibrium configuration of the coupled system. Finally, we extend our conclusions to the case of a single wall interacting with two different cavity fields through a generalization of the effective Hamiltonian, allowing one to include the effect of vacuum radiation pressure on both sides of the movable wall. We also find that at second order the two cavities act independently, while we infer that they influence each other starting from fourth order in the coupling.

This paper is organized as follows: in Sec. \ref{sec:The system} we describe the system under scrutiny, in particular the Hamiltonian formulation which allows us to model the interaction between the mechanical degrees of freedom of the movable mirror and the field in a cavity. In Sec. \ref{sec:Stationary configuration: dressed state} we review the dressed equilibrium configuration of the system giving the energy shifts for each Hamiltonian term at the second order in the coupling constant. Then, in Sec. \ref{sec:Dynamical dressing process: initial bare state} we investigate the dynamical dressing process of the system and find the local dynamical energy shift, thus investigating in detail the dynamical dressing process of the wall. Finally, Sec. \ref{sec:Two cavities scenario} extends our previous analysis by including the effect of the vacuum field outside the cavity; specifically, we obtain a Hamiltonian that takes into account also the effect of the external radiation pressure and calculate the local dynamical energy shift of the system at second order. Sec. \ref{sec:Conclusions} is devoted to our conclusive remarks.

\section{\label{sec:The system}The model}

\begin{figure}
\centering
\includegraphics[scale=0.35]{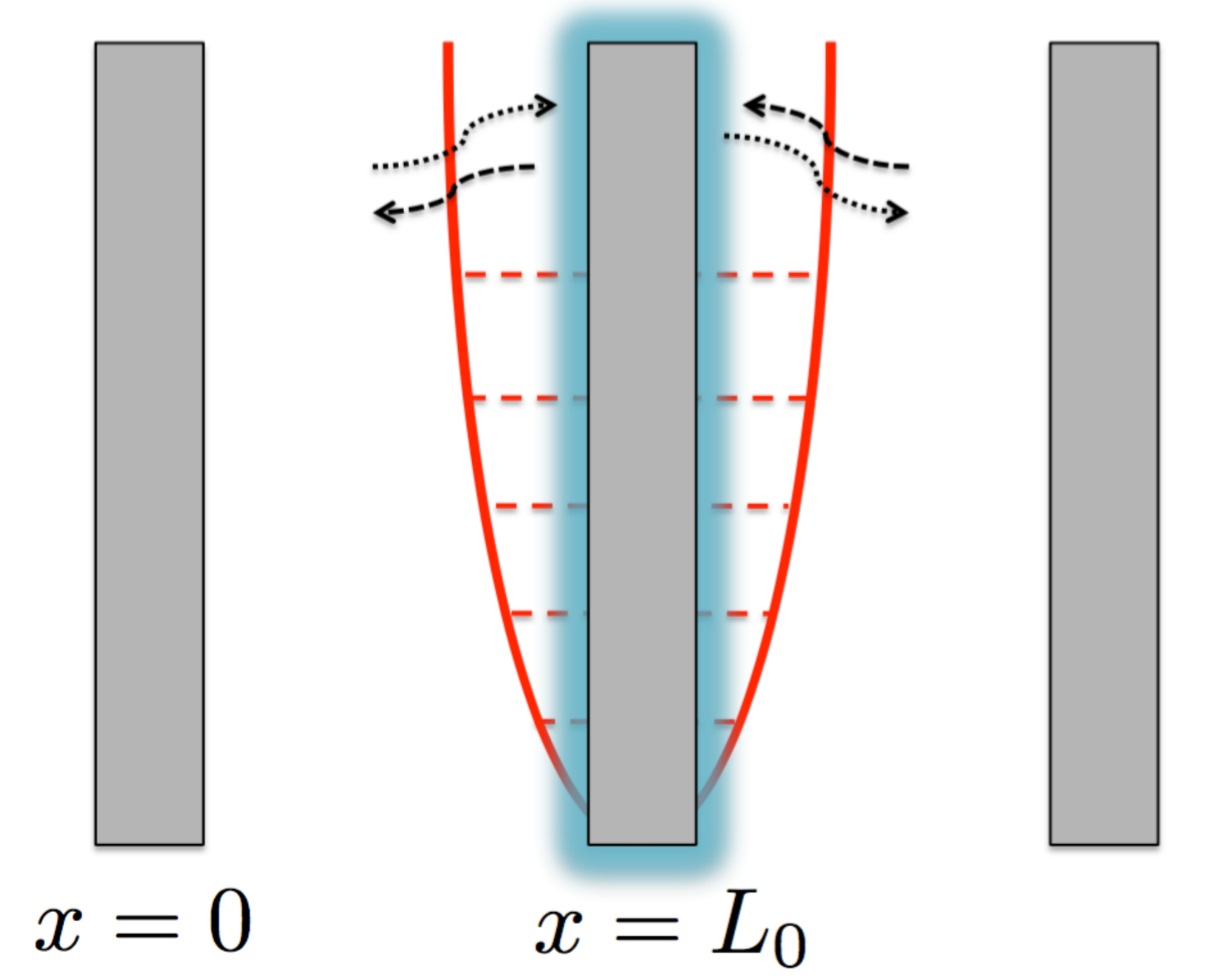}
\caption{Pictorial description of the system under scrutiny. A movable wall, described quantum mechanically, is bounded by a harmonic potential of frequency $\omega_0$ (represented by the red well) around its equilibrium position $L_0$. The wall can interact with two different sets of cavity modes, at its left and right sides. Because of the interaction with the field vacuum fluctuations, the state of the wall starts its time-dependent self-dressing process: the two objects interchange virtual particles (black-dotted arrows).}\label{Fig1}
\end{figure}

Let us consider a one-dimensional massless scalar field inside a cavity formed by two perfectly reflecting mirrors at zero temperature, as depicted in Fig.\ref{Fig1} (the fixed wall shown at the right side of the figure will be introduced later on). The mirror at the left side is fixed at position $x=0$, while the other of mass $M$ is free to move and it is bounded by a harmonic potential of frequency $\wo$ to its equilibrium position $L_0$. The mechanical degrees of freedom of the moving mirror are treated quantum mechanically and included in the overall system dynamics. The effective nonrelativistic Hamiltonian describing our one-dimensional coupled mirror-field system is $\mathcal{H}=\mathcal{H}_0+\mathcal{H}_{i}$, where
\begin{equation}\label{H0}\begin{split}
  \mathcal{H}_{0}&=\mathcal{H}_f+\mathcal{H}_m\\
  &=\hbar\sum_{k}\omega_k\adk\ak+\hbar\wo b^{\dag}b
\end{split}\end{equation}
is the unperturbed Hamiltonian with $\mathcal{H}_f$ and $\mathcal{H}_m$ the field and mirror Hamiltonians, respectively. $b$ ($b^\dagger$) is the annihilation (creation) operator of the mechanical degrees of freedom, while $\ak$ ($\adk$) is the annihilation (creation) operator for the field mode $k$. We impose Dirichlet boundary conditions on the field operator at the positions of the walls; the field modes are relative to the equilibrium position $L_0$ of the moving mirror, and thus the possible wave numbers are $k_j=j\pi /L_0$, with $j$ an integer number. We remark that, since our model is one dimensional, the normalization of the scalar field operator we have used is such that the field energy density inside the cavity has the dimension of energy per unit length, and its integral over the cavity length yields the field energy.
The effective interaction Hamiltonian, describing the interaction between mirror and field and an effective interaction between different field modes (due to the motion of the wall), is \cite{Law1995}
\begin{equation}\label{Hint}
  \mathcal{H}_{i}=-\skj\Ckj(b+\bd)\left[\ak\aj+\adj\ak+\adk\aj+\adk\adj\right] ,
\end{equation}
where
\begin{equation}\label{Ckj}
  \Ckj=(-1)^{k+j}\left( \frac{\hbar}{2}\right) ^{\frac{3}{2}}\frac{1}{L_0\sqrt{M}}\sqrt{\frac{\wk\wj}{\wo}}
\end{equation}
is the coupling constant, while $k$ and $j$ are integer numbers specifying the field modes relative to the equilibrium position $L_0$ of the wall. For the moment, we do not consider the field modes outside the cavity; we will include them in Sec. \ref{sec:Two cavities scenario}. We wish to point out that in our Hamiltonian $\mathcal{H}$ we have taken the zero of the energy such that it includes the static Casimir energy between the two walls $E_{\rm Cas}=- \pi \hbar c/(24L_0)$, and that all energy corrections we are going to discuss are meant as corrections to this value of the energy.

\section{\label{sec:Stationary configuration: dressed state}Stationary configuration: dressed state}

We are now interested in the ground state of our system. The unperturbed ground state is the \textit{bare} state $|\Psi_b\rangle=|\{0_{p}\},0\rangle$, where $\ket{\{0_{p}\}}$ represents the vacuum state for all cavity modes and $\ket{0}$ the vacuum state of the mirror's mechanical excitations. This state is not an eigenstate of the total Hamiltonian because of the mirror-field interaction. The true ground state of the system can then be found applying stationary perturbation theory to the interaction Hamiltonian $(\ref{Hint})$. At the lowest significant order in the coupling constant, it reads \cite{Butera2013}
\begin{equation}\label{ground-Dressed}
  |\Psi_d\rangle=\ket{\Psi_b}+\sum_{jk}\mathcal{D}_{jk} |\{1_{j},1_{k}\},1\rangle  ,
\end{equation}
where
\begin{equation}\label{Dkj}
  \mathcal{D}_{jk}=(-1)^{j+k}\frac{1}{L_{0}}\sqrt{\frac{\hbar\wj\wk}{8M\wo}}\frac{1}{(\wo+\wk+\wj)}.
\end{equation}
The state $|\Psi_d\rangle$ is a \textit{dressed} state of the field-mirror system. It corresponds to the equilibrium configuration of the system where both field and mirror contain virtual excitations. In particular, the elements of the state in \eqref{ground-Dressed} in curly brackets indicate a pair of virtual excitations of the field, while the other element indicates one excitation of the wall (phonon). As discussed in Ref. \cite{Butera2013}, the fact that the state is dressed by a pair of photons is in analogy with the dynamical Casimir effect, where the production of real photons from the vacuum occurs in pairs \cite{Moore70,Dodonov10}. However, in the case of the DCE, the photons emitted are real quanta, emitted in an energy-conserving process and they are produced by a nonadiabatic change of the mirror position; their frequency is determined by the oscillation frequency of the wall. On the contrary, in the present case all frequencies of virtual field excitations are present. It is also worth mentioning that, in the present case, the dressing effect is more relevant for low mass and/or low binding frequency of the movable wall, i.e. when the constant $D_{jk}$ is larger.

The second-order energy shift of the system due to the interaction can be evaluated as follows \cite{Butera2013}
\begin{eqnarray}\label{E2-stat}
  \mathcal{E}_s^{(2)}&=& -\skj\frac{\hbar^2}{4L_0^2M}\frac{\wk\wj}{\wo}\frac{1}{(\wo+\wk+\wj)} \nonumber \\
  &=& \frac 12 {\bra{\Psi_d}\mathcal{H}_{i}\ket{\Psi_d}},
\end{eqnarray}
where $\mathcal{H}_{i}$ is the time-independent interaction Hamiltonian. This energy shift yields a change of the Casimir energy of the system with respect to the configuration in which the walls are fixed. The last equality in \eqref{E2-stat} will be exploited in the next section for our generalization to the dynamical case.\\
Equation \eqref{E2-stat} needs a regularization to cure ultraviolet divergences, i.e. an upper cutoff frequency in the sum over the field modes. An upper cutoff is motivated by the fact that a real mirror becomes transparent at frequencies larger than its plasma frequency, and thus its interaction with the field is strongly suppressed above such a frequency. The absolute value of the energy shift \eqref{E2-stat} increases when the mass and oscillation frequency of the mirror are decreased. This offers the possibility to probe this effect due to the wall's quantum movement (position fluctuations) in the near future since in modern quantum optomechanics experiments very small masses are commonly obtained \cite{aspelmeyer2014}. Also, today-state-of-art experiments probing the Casimir force, the precision is around a few percent \cite{Lamoreaux2005,Woods2016}.\\
As it has already been highlighted in Ref. \cite{Butera2013}, this second order energy shift can be physically interpreted in terms of emitted photons and mirror oscillation energy (virtual excitations), as well as mirror-field interaction energy. Indeed, the stationary energy shifts of each Hamiltonian term are
\begin{equation}\label{H0-Hint-dressed}\begin{split}
\mathcal{E}_{f,s}^{(2)}&=\bra{\Psi_d}\mathcal{H}_f\ket{\Psi_d}\\
&=\skj\frac{\hbar^2}{2L_0^2M}\frac{\wk^2\wj}{\wo(\wo+\wk+\wj)^2},\\
\mathcal{E}_{m,s}^{(2)}&=\bra{\Psi_d}\mathcal{H}_m\ket{\Psi_d}\\
&=\sum_j\frac{\hbar^2}{4L_0^2M}\frac{\wk\wj}{(\wo+\wk+\wj)^2},\\
\mathcal{E}_{i,s}^{(2)}&=\bra{\Psi_d}\mathcal{H}_{i}\ket{\Psi_d}=2\;\mathcal{E}_s^{(2)}.
\end{split}\end{equation}
By using Eqs. \eqref{E2-stat} and \eqref{H0-Hint-dressed} it is then possible to verify that the sum of the energy shifts of field and mirror is related to the interaction energy shift
(the total energy shift is
$ \mathcal{E}_s^{(2)}=\mathcal{E}_{f,s}^{(2)}+\mathcal{E}_{m,s}^{(2)}+\mathcal{E}_{i,s}^{(2)}$)
by the following relations
\begin{equation}\label{H0-Hint-dressed-2}\begin{split}
\mathcal{E}_{f,s}^{(2)}+\mathcal{E}_{m,s}^{(2)}=-\frac{1}{2}\mathcal{E}_{i,s}^{(2)}=-\mathcal{E}_{s}^{(2)}.
\end{split}\end{equation}
In particular, the energy shift of the system receives a positive contribution from the energy stored in both field and oscillating mirror, and a negative contribution from the mirror-field interaction, which is twice the stationary energy shift of the system. These considerations will be important for the dynamical situation we will discuss in the next section, where we indeed generalize the relation between the energy shift and the average value of the interaction Hamiltonian on the dressed ground state (Eqs. \eqref{E2-stat} and \eqref{H0-Hint-dressed-2}) to the time-dependent case we are going to investigate.

\section{\label{sec:Dynamical dressing process: initial bare state}Dynamical dressing of the movable mirror: bare initial state}

We now assume both field and mirror initially ($t=0$) prepared in their bare ground state $|\Psi_b\rangle=|\{0_{p}\},0\rangle$. Also, at $t=0$ we turn on the interaction: field and mirror then start to interact through the Hamiltonian \eqref{Hint} and exchange virtual excitations, since we are starting from a nonequilibrium configuration. This process, which takes place after the initial time $t=0$ and eventually brings the system to an equilibrium configuration, is the \textit{dynamical dressing process} we wish to investigate. The final configuration of the system is a \textit{dressed state}, which, however, does not coincide with the time-independent state $\ket{\Psi_d}$ obtained in Eq.\eqref{ground-Dressed} by a stationary approach.

In order to describe this situation, we follow the ideas developed in the context of the dynamical Casimir-Polder interaction between an atom and a fixed wall in \cite{VP08,MVP10,Armata16}, consisting in extending to the time-dependent case the results given in the previous section. We evaluate in the Heisenberg representation the time-dependent interaction energy at second order in the coupling, by using
\begin{equation}\label{Eshift-dyn}
\mathcal{E}^{(2)}_{\rm loc,d}(t)=\frac{\bra{\Psi_b}\mathcal{H}^{(2)}_{i}(t)\ket{\Psi_b}}{2},
\end{equation}
that is the natural way of generalizing Eqs. \eqref{E2-stat} (last line) and \eqref{H0-Hint-dressed-2}, where $\mathcal{H}^{(2)}_{i}(t)$ is the interaction Hamiltonian \eqref{Hint} in the Heisenberg representation. The energy shift in Eq.\eqref{Eshift-dyn} is referred to as a \textit{local} one since, contrary to the stationary case, it describes the \emph{local} interaction between the plate and the field, and in general does not coincide (in particular at very large times) with the overall energy shift of the system. This is generally due to the fact that the radiation emitted during the self-dressing process of the system propagates to large distances \cite{Armata16,PPP95,CV99,POP01}. On the other hand, we also wish to stress that the overall energy shift of the system is time independent due to the unitary time evolution. Later on in this section we shall discuss this point in more detail.

To calculate $\mathcal{H}^{(2)}_{i}(t)$, we first solve the Heisenberg equations for the field and plate operators by perturbation theory. We thus obtain the time-dependent field and plate annihilation and creation operators as a series expansion in the coupling constant, $a_k(t)=a_k^{(0)}(t)+a_k^{(1)}(t)+...+a_k^{(i)}(t)+...$ and $b(t)=b^{(0)}(t)+b^{(1)}(t)+...+b^{(i)}(t)+...$, where $a_k^{(i)}(t)$ and $b^{(i)}(t)$ are proportional to the $i$-th power of $\mathcal{C}_{kj}$. At the lowest significant order, the Heisenberg equations for the mirror and field operators give
\begin{equation}\label{FirstCorr-1}\begin{split}
b^{(0)}(t)=b\ewom,\quad a_k^{(0)}(t)=\ak \ewkm ,
\end{split}\end{equation}
\begin{equation}\label{FirstCorr-1a}\begin{split}
&b^{(1)}(t)=\frac{i}{\hbar}\skj\mathcal{C}_{kj}\ewom\\
&\times[\ak\aj \mathcal{F}(\wo-\wk-\wj)+\adj\ak \mathcal{F}(\wo-\wk+\wj)\\
&\hspace{0.2cm}+\adk\aj \mathcal{F}(\wk-\wj+\wo)+\adk\adj \mathcal{F}(\wk+\wj+\wo)],
\end{split}\end{equation}
\begin{equation}\label{FirstCorr-1b}\begin{split}
&a^{(1)}_k(t)=\frac{2i}{\hbar}\sum_j\mathcal{C}_{kj}\ewkm\\
&\times[\aj b\mathcal{F}(\wk-\wj-\wo)+\aj\bd \mathcal{F}(\wk-\wj+\wo)\\
&\hspace{0.2cm}+\adj b\mathcal{F}(\wk+\wj-\wo)+\adj\bd \mathcal{F}(\wk+\wj+\wo)],
\end{split}\end{equation}
where
\begin{equation}\label{F}
\mathcal{F}(\omega)=\frac{e^{i\omega t}-1}{i\omega}
\end{equation}
(the operators without an explicit time dependence are at time $t=0$). We use \eqref{FirstCorr-1}, \eqref{FirstCorr-1a}, \eqref{FirstCorr-1b} in the expression of  $\mathcal{H}^{(2)}_{i}(t)$ as given by \eqref{Hint}, written in the Heisenberg representation, maintaining only terms up to the second order in the coupling. We obtain the expression of the interaction Hamiltonian $\mathcal{H}^{(2)}_{i}(t)$ as the sum of two different contributions, $\mathcal{H}^{(2)}_{i}(t)=\mathcal{H}^{(2)}_{i,a}(t)+\mathcal{H}^{(2)}_{i,b}(t)$, where the first term is given by the substitution of the zeroth order mirror operator $b^{(0)}(t)$ and the first-order field operator $a^{(1)}_k(t)$, while the second by the substitution of the correction $b^{(1)}(t)$ at first order for the mirror operator  and at zeroth order for the field operator $a^{(0)}_k(t)$. We have
\begin{equation}\label{Hint-a}\begin{split}
 &\mathcal{H}^{(2)}_{i,a}(t)=-\skj\Ckj\left[b^{(0)}(t)+b^{\dagger^{(0)}}(t)\right]\\
 &\hspace{0.5cm}\times\big[a_{k}^{(1)}(t)a_{j}^{(0)}(t)+a_{k}^{\dagger^{(1)}}(t) a_{j}^{\dagger^{(0)}}(t)+a_{k}^{\dagger^{(1)}}(t) a_{j}^{(0)}(t)\\
 &\hspace{3cm}+a_{j}^{\dagger^{(1)}}(t) a_{k}^{(0)}(t)+\left\lbrace(1)\leftrightarrow (0)\right\rbrace\big] ,
\end{split}\end{equation}
\begin{equation}\label{Hint-b}\begin{split}
& \mathcal{H}^{(2)}_{i,b}(t)=-\skj\Ckj\left[b^{(1)}(t)+b^{\dagger^{(1)}}(t)\right]\\
 &\hspace{0.5cm}\times\big[a_{k}^{(0)}(t)a_{j}^{(0)}(t)+a_{k}^{\dagger^{(0)}}(t) a_{j}^{\dagger^{(0)}}(t)\\
 &\hspace{3cm}+a_{k}^{\dagger^{(0)}}(t) a_{j}^{(0)}(t)+a_{j}^{\dagger^{(0)}}(t) a_{k}^{(0)}(t)\big] ,
\end{split}\end{equation}
where in the last term of \eqref{Hint-a} the two indices $(0)$ and $(1)$ are swapped. By substituting the expressions given by \eqref{FirstCorr-1}, \eqref{FirstCorr-1a}, \eqref{FirstCorr-1b} into \eqref{Hint-a} and \eqref{Hint-b}, we finally obtain
\begin{widetext}
\begin{equation}\label{Hint-a-b}\begin{split}
&\mathcal{H}^{(2)}_{i,a}(t)=-\frac{2i}{\hbar}\sum_{kj}\Ckj\left[b\ewom+\bd\ewo\right]\\
&\times\bigg\lbrace
\bigg[\sum_{j'}C_{jj'}\ak[a_{j'}b\mathcal{F}(\wj-\omega_{j'}-\wo)+a_{j'}\bd \mathcal{F}(\wj-\omega_{j'}+\wo)+a_{j'}^{\dagger}b\mathcal{F}(\wj+\omega_{j'}-\wo)+a_{j'}^{\dagger}\bd \mathcal{F}(\wj+\omega_{j'}+\wo)]\\
 &\hspace{0.5cm}+C_{kj'}[a_{j'}b\mathcal{F}(\wk-\omega_{j'}-\wo)+a_{j'}\bd \mathcal{F}(\wk-\omega_{j'}+\wo)+a_{j'}^{\dagger}b\mathcal{F}(\wk+\omega_{j'}-\wo)+a_{j'}^{\dagger}\bd\mathcal{F}(\wk+\omega_{j'}+\wo)]\aj\bigg]   \\
 &\hspace{0.5cm} \times e^{-i(\wj+\wk)t}\\
&+\bigg[\sum_{j'}C_{jj'}\adk[a_{j'}b\mathcal{F}(\wj-\omega_{j'}-\wo)+a_{j'}\bd \mathcal{F}(\wj-\omega_{j'}+\wo)+a_{j'}^{\dagger}b\mathcal{F}(\wj+\omega_{j'}-\wo)+a_{j'}^{\dagger}\bd \mathcal{F}(\wj+\omega_{j'}+\wo)]\\
&\hspace{0.5cm}-C_{kj'}[a_{j'}^{\dagger}b^{\dagger}\mathcal{F}^*(\wk-\omega_{j'}-\wo)+a_{j'}^{\dagger}b \mathcal{F}^*(\wk-\omega_{j'}+\wo)+a_{j'}b^{\dagger}\mathcal{F}^*(\wk+\omega_{j'}-\wo)+a_{j'}b \mathcal{F}^*(\wk+\omega_{j'}+\wo)]\aj\bigg]\\
&\hspace{0.5cm} \times e^{i(\wk-\wj)t}  +H.c.\bigg\rbrace ,
\end{split}\end{equation}
\begin{equation}\label{Hint-a--b}\begin{split}
&\mathcal{H}^{(2)}_{i,b}(t)=-\frac{i}{\hbar}\sum_{kj}\sum_{k'j'}C_{kj}C_{k'j'}\\
&\hspace{0.5cm}\times \bigg\lbrace a_{k'}a_{j'}\left[\ewom \mathcal{F}(\wo-\omega_{k'}-\omega_{j'})-\ewo \mathcal{F}^*(\wo+\omega_{k'}+\omega_{j'})\right]\\
&\hspace{5cm}+a_{k'}^{\dagger}a_{j'}\left[\ewom \mathcal{F}(\wo+\omega_{k'}-\omega_{j'})-\ewo \mathcal{F}^*(\wo-\omega_{k'}+\omega_{j'})\right]-H.c.\bigg\rbrace\\
&\hspace{0.5cm}\times\left[\ak\aj e^{-i(\wk+\wj)t}+\adk\adj e^{i(\wk+\wj)t}+\adk\aj e^{i(\wk-\wj)t}+\adj\ak e^{i(\wj-\wk)t}\right] .
\end{split}\end{equation}
\end{widetext}
We now study the local interaction energy between field and mirror by means of Eq. \eqref{Eshift-dyn}. The average of Eqs. \eqref{Hint-a-b} and \eqref{Hint-a--b} on the initial bare state $\ket{\Psi_b}$, where both field and mirror are in their ground states, yields
\begin{equation}\label{Hint-a-mean}\begin{split}
&\bra{\Psi_b}\mathcal{H}^{(2)}_{i,a}(t)\ket{\Psi_b}=-\frac{2i}{\hbar}\skj\Ckj^2e^{-i\wo t}\\
&\hspace{0.4cm}\times[ e^{-i(\wj+\wk)} \mathcal{F}(\wj+\wk+\wo)\\
&\hspace{3cm}-e^{i(\wk+\wj)t}\mathcal{F}^*(\wk+\wj-\wo)],
\end{split}\end{equation}
\begin{equation}\label{Hint-b-mean}\begin{split}
&\bra{\Psi_b}\mathcal{H}^{(2)}_{i,b}(t)\ket{\Psi_b}=-\frac{2i}{\hbar}\skj\Ckj^2e^{i(\wk+\wj)t}\\
&\hspace{0.4cm}\times\left[\ewom \mathcal{F}(\wo-\wj-\wk)-\ewo \mathcal{F}^*(\wj+\wk+\wo)\right] .
\end{split}\end{equation}
Summing up the results above and using \eqref{Ckj}, we can finally obtain the second-order local dynamical interaction energy
\begin{equation}\label{E2-dyn}\begin{split}
&\mathcal{E}^{(2)}_{\rm loc,d}(t)=-\frac{\hbar^2}{4L_0^2M}\skj\frac{\wk\wj}{\wo} \frac{1}{\wo+\wk+\wj}\\
&\hspace{1cm}\times\left\lbrace 1- \cos\left[\left(\wo+\wk+\wj\right)t\right]\right\rbrace .
\end{split}\end{equation}
\begin{figure}
\centering
\includegraphics[scale=0.50]{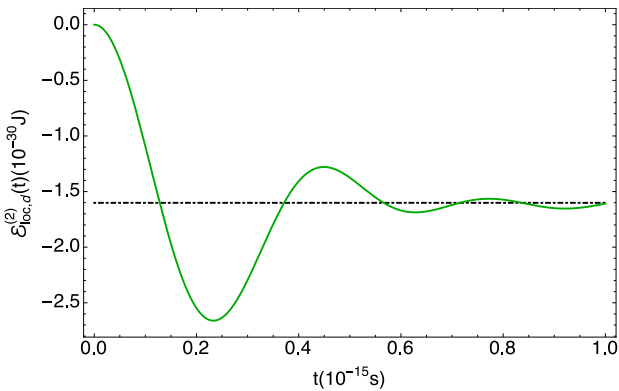}
\caption{The plot (green continuous line) shows the time evolution of the dynamical interaction energy shift of the system in the continuous limit (Eq.\eqref{E2-dyn-continuous}). It shows oscillations around its stationary value ($t \to \infty$) represented by the black dot-dashed line, which also coincides with the overall energy shift in the fully dressed configuration (Eq. \eqref{E2-stat}). The figure shows that there are time intervals where the dynamical interaction energy is larger than its stationary value.The numerical values used for the angular frequency and mass of the movable mirror are respectively $\wo=10^4 \text{$s^{-1}$}$ and $\text{M}=10^{-14}$Kg, while the cutoff frequency has been set to $\omega_{\rm cut}=10^{16}\text{$s^{-1}$}$.}
\label{Fig2}
\end{figure}
Equation \eqref{E2-dyn} describes the \textit{local} interaction energy between the field inside the cavity and the mobile wall. We note that it is zero at $t=0$, when the interaction has not taken place yet. On the other hand, it is worthwhile to study the opposite limit of very large times, since it corresponds to the stationary regime for the system. We first discuss such a limit in the case of a continuous set of modes in the cavity, that is when we bring the cavity length to infinity ($L_0\rightarrow\infty$). In this case we have $\sum_{kj}\rightarrow L_0^2/(2\pi)^2\int_0^{\infty}\int_0^{\infty}dkdk'$, and the energy shift in Eq.\eqref{E2-dyn} can be rewritten as
\begin{equation}\label{E2-dyn-continuous}\begin{split}
\mathcal{E}^{(2)}_{\rm loc,d}(t)&=-\frac{\hbar^2\omega_0^2}{16\pi^2Mc^2}\int_0^{\infty}dx\int_0^{\infty}dx'\\
&\times\frac{xx'}{1+x+x'}\left\lbrace 1- \cos\left[a\left(1+x+x'\right)t\right]\right\rbrace,
\end{split}\end{equation}
where $a=\omega_0t$, and $x=ck/\omega_0$. In the limit $t\rightarrow\infty$, since the cosine function appearing in Eq. \eqref{E2-dyn-continuous} is a rapidly oscillating function, its value averages to zero, and we easily obtain
\begin{equation}\label{E2-dyn-limit}\begin{split}
\mathcal{E}^{(2)}_{loc,d}(t)\longrightarrow\mathcal{E}^{(2)}_{s},
\end{split}\end{equation}
As a consequence, for large times, the local dynamical energy shift coincides with the static result for the overall energy shift of the system (see Eq. \eqref{E2-stat}), as obtained when the state of the system is the fully dressed state \eqref{ground-Dressed}. This confirms that our way for generalizing the second-order energy shift to the dynamical case (Eq. \eqref{Eshift-dyn}) is physically sound. This procedure has already given consistent results for the dynamical Casimir-Polder interaction energy between an initially bare or partially dressed atom and a perfectly conducting wall \cite{VP08,MVP10,Armata16}.

In order to evaluate explicitly Eq. \eqref{E2-dyn-continuous}, we need to introduce an ultraviolet cutoff frequency. As already pointed out, this is necessary from a physical point of view because a real conducting wall is characterized by a plasma frequency, and field cavity modes with a frequency higher than the plasma frequency do not experience the presence of the boundary, and thus do not contribute to the effective wall-field interaction. Figure \ref{Fig2} shows a plot of the second-order energy \eqref{E2-dyn-continuous} as a function of time by numerically solving the integral and using a sharp frequency cutoff. The plot shows that the interaction energy vanishes at $t=0$ and at successive times it shows oscillations around its stationary limit $\mathcal{E}^{(2)}_{s}$. Also, in specific time intervals the absolute value of the dynamical interaction energy can be larger than in the static case, thus leading to an increase of the Casimir energy of the system. This new feature could be an important aspect for observing the dynamical effect.

\begin{figure}[t]
\centering
\subfigure{
        \label{Time_Evol_Field}
     \includegraphics[width=0.42\textwidth]{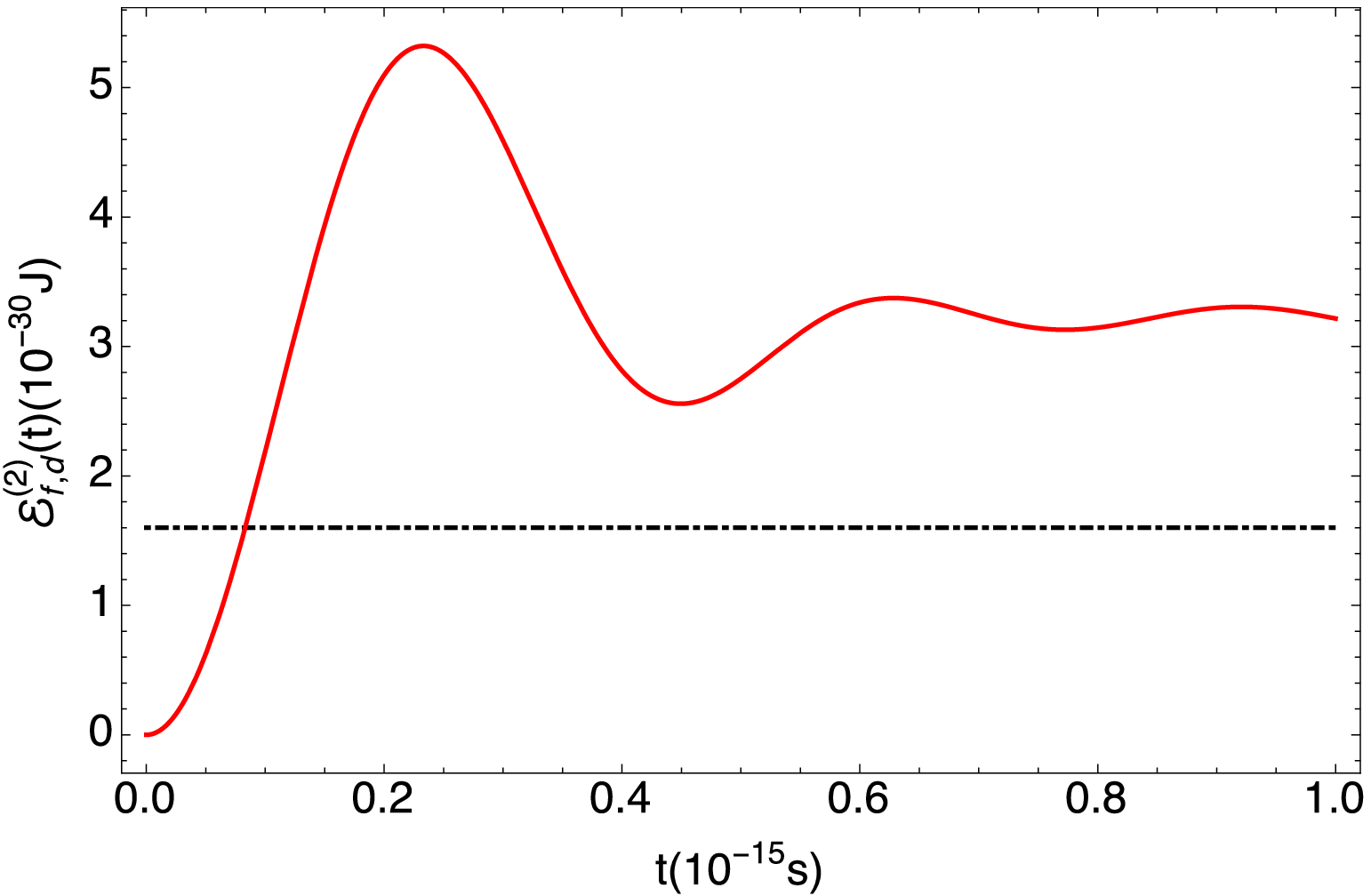} }
\subfigure[]{
        \label{Time_Evol_Mirror}
        \includegraphics[width=0.42\textwidth]{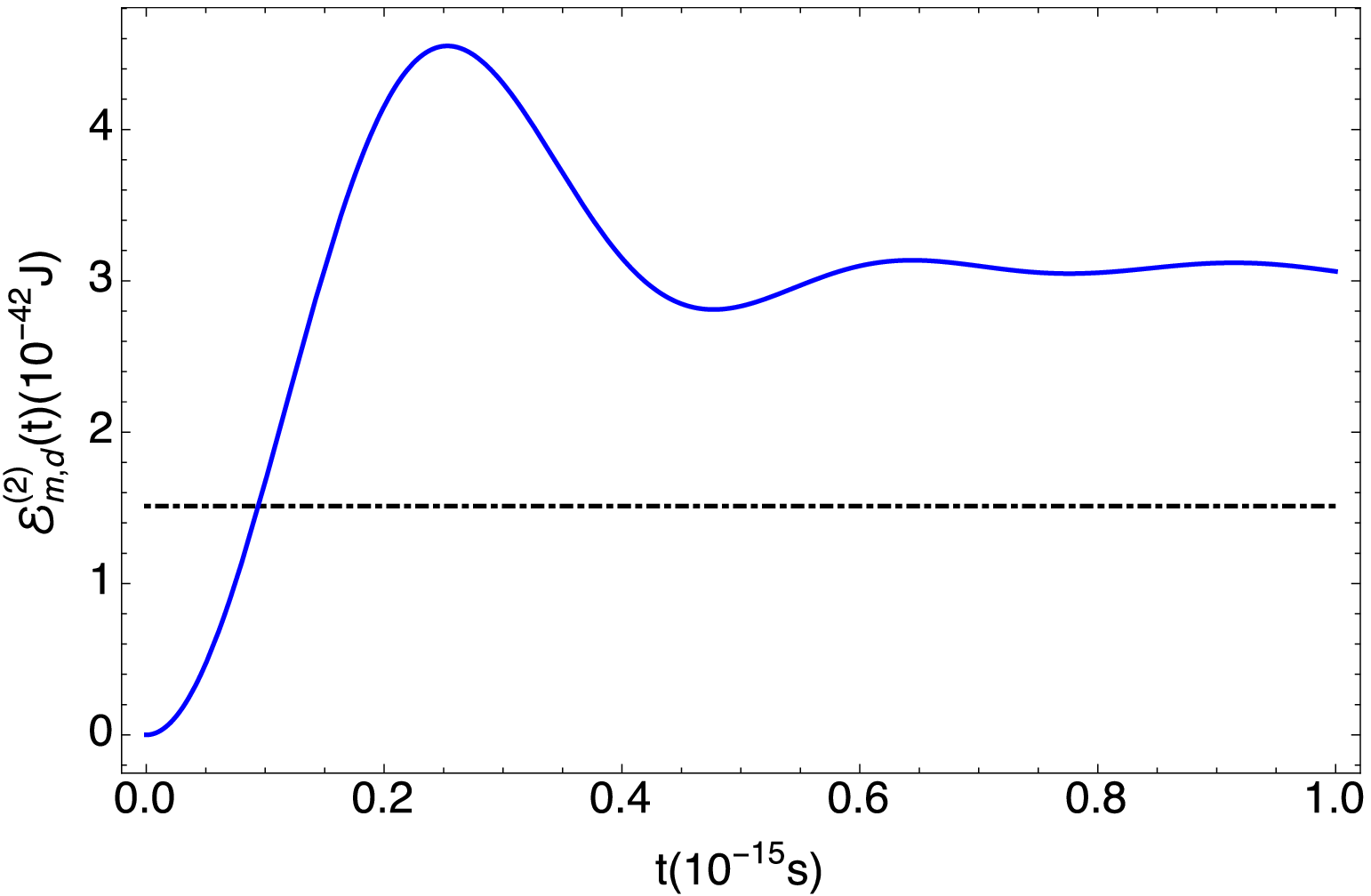} }
\caption{Time evolution of dynamical energy shifts in the continuous limit $L_0\rightarrow\infty$ (Eq.\eqref{Hfm-shift-continuous}): (a) field energy (red continuous line); (b) mirror energy (blue continuous line). Their dynamics shows oscillations which tend to twice the corresponding stationary value, represented by the black dot-dashed line. The numerical values of the parameters are the same as in the plot in Fig. \ref{Fig2}.}
\label{Fig3}
\end{figure}

It is also worth comparing the local dynamical energy shift \eqref{E2-dyn} with the dynamical shifts of the unperturbed field and mirror Hamiltonians, thus separating the single contributions to the overall energy shift. Following the same procedure as before, we substitute the perturbative solutions \eqref{FirstCorr-1}, \eqref{FirstCorr-1a} and \eqref{FirstCorr-1b} for the mirror and field  operators into the expressions of the field and mirror Hamiltonians. After averaging on the initial bare state of the system $\ket{\Psi_b}$, we obtain
\begin{equation}\label{Hf-shift}\begin{split}
\mathcal{E}_{f,d}^{(2)}(t)&=\bra{\Psi_b}\mathcal{H}_{f}^{(2)}(t)\ket{\Psi_b}\\
&=\frac{\hbar^2}{L_0^2M}\skj\frac{\wk^2\wj}{\wo}\frac{1}{(\wo+\wk+\wj)^2}\\
&\hspace{2cm}\times\left\lbrace 1-\cos[(\wo+\wk+\wj)t]\right\rbrace ,
\end{split}\end{equation}
\begin{equation}\label{Hm-shift}\begin{split}
\mathcal{E}_{m,d}^{(2)}(t)&=\bra{\Psi_b}\mathcal{H}_{m}^{(2)}(t)\ket{\Psi_b}\\
&=\frac{\hbar^2}{2L_0^2M}\skj\frac{\wk\wj}{(\wo+\wk+\wj)^2}\\
&\hspace{2cm}\times\left\lbrace 1-\cos[(\wo+\wk+\wj)t]\right\rbrace .
\end{split}\end{equation}
We can immediately verify that
\begin{equation}\label{H-tot-dyn}\begin{split}
\bra{\Psi_b}\mathcal{H}^{(2)}(t)\ket{\Psi_b}=0,
\end{split}\end{equation}
as expected since the system undergoes a unitary evolution from the nonequilibrium initial state $|\Psi_b\rangle$, and thus the total average energy of the system is constant in time. We now discuss the limit for large times ($t\rightarrow\infty$) of the field and mirror energy shifts. We again study such limit in the continuous limit for the cavity modes. Equations \eqref{Hf-shift} and \eqref{Hm-shift} thus become
\begin{equation}\label{Hfm-shift-continuous}\begin{split}
\mathcal{E}_{f,d}^{(2)}(t)&=\frac{\hbar^2\omega_0^2}{4\pi^2Mc^2}\int_0^{\infty}dx\int_0^{\infty}dx'\\
&\times\frac{x^2x'}{(1+x+x')^2}\left\lbrace 1- \cos\left[a\left(1+x+x'\right)t\right]\right\rbrace ,\\
\mathcal{E}_{m,d}^{(2)}(t)&=\frac{\hbar^2\omega_0^2}{8\pi^2Mc^2}\int_0^{\infty}dx\int_0^{\infty}dx'\\
&\times\frac{xx'}{(1+x+x')^2}\left\lbrace 1- \cos\left[a\left(1+x+x'\right)t\right]\right\rbrace,
\end{split}\end{equation}
and for $t\rightarrow\infty$ we get
\begin{equation}\label{Limits-fm}\begin{split}
\mathcal{E}_{f,d}^{(2)}(t)&\longrightarrow 2\mathcal{E}^{(2)}_{f,s} ,\\
\mathcal{E}_{m,d}^{(2)}(t)&\longrightarrow 2\mathcal{E}^{(2)}_{m,s}.
\end{split}\end{equation}
From these limits we can deduce that both dynamical energies stored in the field and in the mirror tend to twice their stationary values. This behavior is shown in the two plots of Fig. \ref{Fig3} where the time evolution of the dynamical field and mirror energy shifts are plotted. On the other hand, using Eqs. \eqref{Eshift-dyn} and \eqref{E2-dyn-continuous} we find that at large times the energy stored in the interaction Hamiltonian coincides with its stationary value which is equal to twice the total energy shift of the system in the equilibrium (fully dressed) configuration
\begin{equation}\label{Limits-int}\begin{split}
\mathcal{E}^{(2)}_{i,d}&\longrightarrow \mathcal{E}^{(2)}_{i,s}=2\mathcal{E}^{(2)}_{s}.
\end{split}\end{equation}
This means that during the dressing process, since the energy stored in the interaction will approach to its stationary value, the field and the mirror must raise their energy in order to preserve the energy conservation expressed by Eq. \eqref{H-tot-dyn}. Nevertheless, the additional energy stored in field and oscillator takes into account the difference between the two (dressed and bare, respectively) initial configurations for the stationary and the dynamical regimes. This shows a rather subtle aspect of the approach to equilibrium of our system: a \emph{local} quantity such as the mirror-field interaction energy (localized at the mirror's position) tends to its stationary value, while \emph{global} quantities such as the field energy (which includes contributions from field emitted during the self-dressing \cite{PPP95,CV99,POP01} and propagating at a very large distance) do not approach their equilibrium values (as obtained by the stationary approach). A similar behaviour was obtained for the dynamical dressing, and related Casimir-Polder energies, of an atom near a conducting wall \cite{Armata16}.

\begin{figure}[t]
\centering
\includegraphics[scale=0.30]{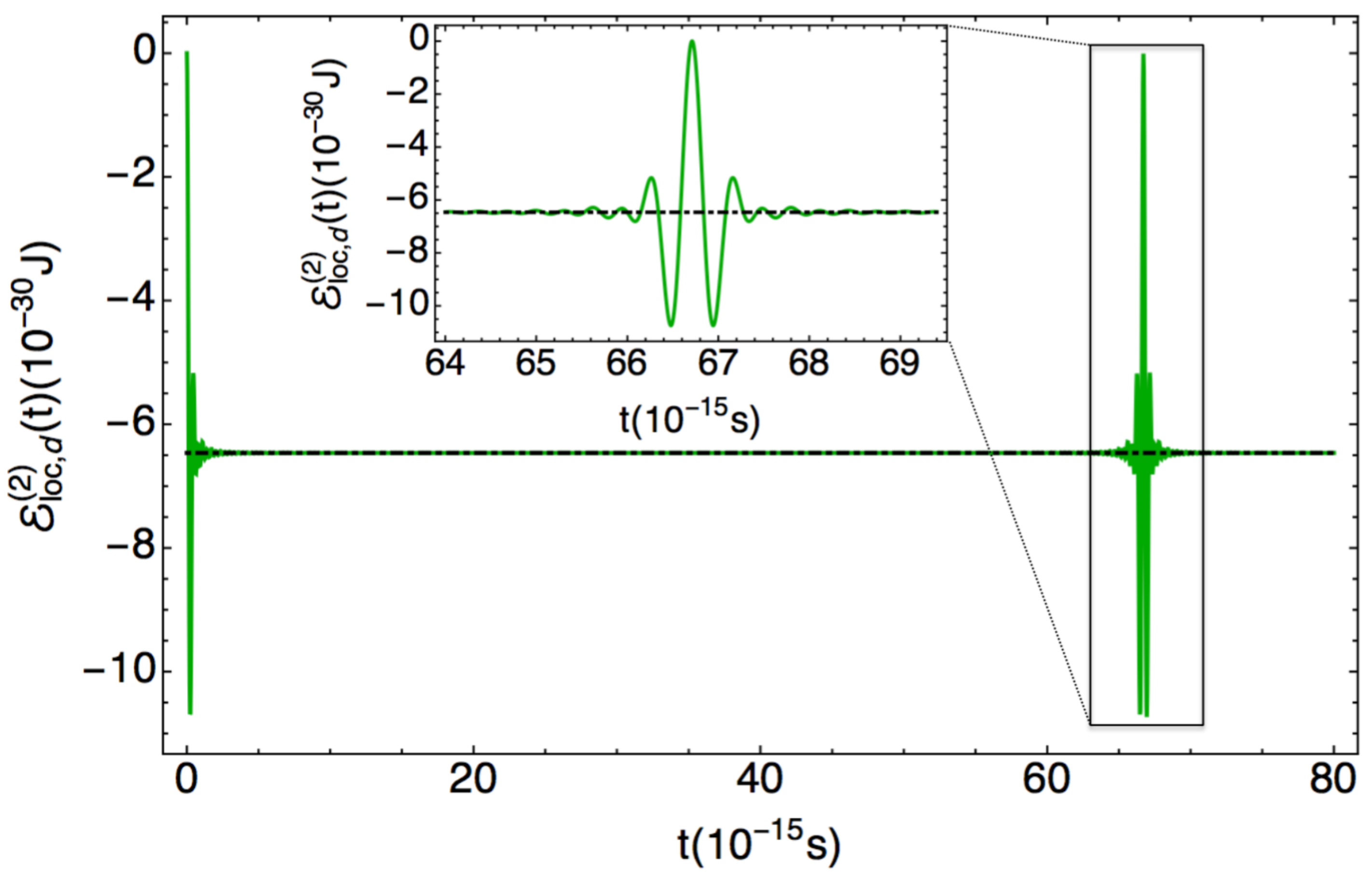}
\caption{The plot (green continuous line) shows the time evolution of the dynamical interaction energy in the case of a finite value for the distance between the walls. After the oscillations around its stationary value (represented by the black dot-dashed line), the interaction energy approaches its stationary value, and then it reappears again. The revivals occur for integer multiples of the round-trip time $\bar{t}=2L_0/c$. We have used $L_0=10^{-5}$, and the other numerical values are as in Fig.\ref{Fig2}.}
\label{Fig5}
\end{figure}

In all the previous discussions the limit for large times has been studied for the case in which $L_0\rightarrow\infty$, which corresponds to the dynamical dressing of a single wall in the vacuum of the electromagnetic field. We finally want to discuss such limit directly for Eqs. \eqref{E2-dyn}, \eqref{Hf-shift}, \eqref{Hm-shift}, that is when we have a discrete set of modes inside the cavity, or in other words, when the distance between the walls is finite. Actually, the validity of the limits in \eqref{E2-dyn-limit}, \eqref{Limits-fm}, and \eqref{Limits-int} can safely be extended to a finite cavity length, as soon as we consider times shorter than the round-trip time of a photon inside the cavity, i.e. $\bar{t}=2L_0/c$. Indeed, for integer multiples of $\bar{t}$,
Eq. \eqref{E2-dyn} shows a revival, as shown in Fig.\ref{Fig5} (the same happens for \eqref{Hf-shift} and \eqref{Hm-shift}). This is a peculiarity of the system under consideration, which does not find a correspondence in the dynamical study of the local energy shift between an atom and a conducting wall. In fact, in that case, the radiation emitted by the atom, which is in part reflected by the wall, propagates to infinity and no longer participates in the dynamical dressing of the wall. On the contrary, in the present case, the radiation emitted during the dynamical dressing at the proximity of the movable wall propagates along the cavity and then is reflected by the fixed mirror, thus yielding the revivals we find. However, when we take the continuous limit $L_0\rightarrow\infty$, we let this energy propagate to infinity, thereby ensuring that it does not contribute to the dressing process for large times.

\section{\label{sec:Two cavities scenario}Two cavities scenario}

The Hamiltonian \eqref{Hint} includes the quantum field in the space between the two plates, while the field in the regions external to the plates was neglected. This is a common scenario in quantum optics, for example, where usually a single field mode is coupled with the mobile mirror \cite{Law1995,aspelmeyer2014}. In other cases, the inclusion of the field external to the cavity might be relevant too. In this section we intend to extend our model by including the effect of the vacuum field fluctuations on the right (external) side of the movable mirror on the dynamical dressing process of the system we have considered. This actually means to consider two distinct cavities which share the same movable mirror (see Fig. 1).  Since the movable mirror is assumed to be perfectly reflective for both cavities, the effective Hamiltonian \eqref{Hint} can easily be generalized to the present case. The unperturbed and interaction Hamiltonians can now be rewritten as
\begin{equation}\label{H0-Hint-2}\begin{split}
  \tilde{\mathcal{H}}_{0}&=\hbar\wo b^{\dag}b+\hbar\sum_{k}\sum_{l=1}^{2}[\omega_{k,l}\adkl\akl] ,\\
 \tilde{\mathcal{H}}_{i}&=-(b+\bd)\sum_{kj}\sum_{l=1}^{2}\Ckjl[\akl\ajl+\adjl\ak\\
 &\hspace{4cm}+\adkl\ajl
 +\adkl\adjl],
\end{split}\end{equation}
where $\aku$ ($\akd$) are the field operators for the two cavities; $\omega_{k,i}=ck\pi/L_0^i$ and $L_0^i$ ($i=1,2$), respectively, are the generic equilibrium frequencies and lengths of the two cavities. Also, the new coupling constants $\Ckju$ and $\Ckjd$ are accordingly modified, though having the same form as in Eq. \eqref{Ckj}. However, if we take the value of $\Ckju$ as that given by \eqref{Ckj}, the value of $\Ckjd$ must be taken with the opposite sign. In fact, in the expansion in the mirror's displacement around its equilibrium position leading to \eqref{Hint} and \eqref{Ckj} in \cite{Law1995}, the linear term has a different sign depending on the direction of the wall's displacement. It is also worth mentioning that the generalization in Eq.\eqref{H0-Hint-2} is possible only because the Hamiltonian \eqref{Hint} is an \textit{effective} Hamiltonian, and thus the interaction between the cavity modes and the mechanical degrees of freedom is encoded in its operatorial form and not in the physical parameters such as the cavity length or frequency, which is indeed related to the equilibrium position of the moving wall. For this reason, the only way by which the fields inside the two cavities can interact with each other is through the mirror's displacement operator. Therefore our Hamiltonian \eqref{H0-Hint-2} can safely be used under the assumption of a perfectly conducting mobile boundary, allowing an interaction between the field in the two cavities only through the mirror's movement. The resulting equations of motion for mirror and field operators at the zeroth and first order in the coupling constants, obtained similarly to the single-cavity case of the previous section, are then
\begin{equation}\label{Zero-2cavity}
b^{(0)}(t)=b\ewom ,\hspace{0.1cm}
\akl^{(0)}(t)=\akl e^{-i\omega_{k,l}t},
\end{equation}
\begin{widetext}
\begin{equation}\label{FirstCorr-2cavity}\begin{split}
&b^{(1)}(t)=\frac{i}{\hbar}\ewom\skj \sum_{l=1}^2 \mathcal{C}_{kj}^l\big[\akl\ajl\mathcal{F}(\wo-\omega_{k,l}-\omega_{j,l})+a^{\dag}_{j,l} a_{k,l} \mathcal{F}(\wo-\omega_{k,l} +\omega_{j,l}) \\
&\hspace{5cm}+\adkl\ajl \mathcal{F}(\wo+\omega_{k,l}-\omega_{j,l})+\adkl\adjl \mathcal{F}(\omega_{k,l}+\omega_{j,l}+\wo)\big],\\
&\akl^{(1)}(t)=\frac{2i}{\hbar}\sum_j\mathcal{C}_{jk}^le^{-i\omega_{k,l}t}\big[\ajl b\mathcal{F}(\omega_{k,l}-\omega_{j,l}-\wo)+\ajl\bd \mathcal{F}(\omega_{k,l}-\omega_{j,l}+\wo)\\
&\hspace{5cm}+\adjl b\mathcal{F}(\omega_{k,l}+\omega_{j,l}-\wo)+\adjl\bd \mathcal{F}(\omega_{k,l}+\omega_{j,l}+\wo)\big] ,
\end{split}\end{equation}
where the operators without an explicit time dependence are at $t=0$. Inspection of \eqref{FirstCorr-2cavity} shows that, at the second order in the coupling constants, the field operators are independent from each other, whereas the mirror operator depends on the field operators relative to both cavities, according to our previous physical considerations.

We now explore how the mirror's dressing process is modified by the interaction with the two different cavity fields. Following the same procedure of Sec \ref{sec:Dynamical dressing process: initial bare state}, we consider an initial bare state where both mirror and cavity fields are in the vacuum state and insert Eqs. \eqref{Zero-2cavity}, \eqref{FirstCorr-2cavity}, into Eq. \eqref{Eshift-dyn}. After taking the expectation value on the initial bare state, we obtain the local dynamical energy shift of the system in the following form
\begin{equation}\label{E2-dyn-2}\begin{split}
\tilde{\mathcal{E}}^{(2)}_{\rm loc,d}(t)&=\frac{\bra{\Psi_b}\tilde{\mathcal{H}}^{(2)}_{i}(t)\ket{\Psi_b}}{2}\\
&=-\frac{\hbar^2}{4M}\skj\sum_{l=1}^{2}\frac{1}{(L_0^l)^2}\frac{\omega_{k,l}\omega_{j,l}}{\wo} \frac{1}{\wo+\omega_{k,l}+\omega_{j,l}}\left\lbrace 1- \cos\left[(\wo+\omega_{k,l}+\omega_{j,l})t\right]\right\rbrace.
\end{split}\end{equation}
\end{widetext}
Assuming for simplicity that the two cavities have the same equilibrium length $L_0$, we obtain
\begin{equation}\label{E2-dyn-2-bis}\begin{split}
\tilde{\mathcal{E}}^{(2)}_{\rm loc,d}(t)=2\mathcal{E}^{(2)}_{\rm loc,d}(t),
\end{split}\end{equation}
where $\mathcal{E}^{(2)}_{\rm loc,d}(t)$ was obtained in \eqref{E2-dyn} for the single-cavity case.
Thus the local energy shift is twice the local energy shift \eqref{E2-dyn} obtained in the case of a movable mirror interacting with a single cavity field. This result is important also in the case one is interested in studying the dressing process of a single wall in the presence of the electromagnetic vacuum. Indeed, in order to investigate this case we only need to consider the limit $L_0\rightarrow\infty$ for both cavities. Moreover, Eq. \eqref{E2-dyn-2-bis} shows that the wall's dressing process with the two semispaces is independent, being the interaction energy indeed equal to twice the value obtained in the case of a single cavity. This happens even though the two semispaces can interact with each other through the mirror movement. However, we wish to stress that this holds only at the second order in the coupling constant. At the next nonvanishing order, specifically at fourth order in the coupling constant, we expect that the independence of the two cavity fields will be lost and that the dressing processes in the two semispaces will result to be correlated, and this should have remarkable effects on the Casimir force between two (fixed) walls when a perfectly conducting mobile wall is inserted between them. We will address this aspect, and its relevance for Casimir forces, in a subsequent work.

\section{\label{sec:Conclusions}Conclusions}
In this paper we have considered the dynamical interaction energy between a one-dimensional massless scalar field and a movable mirror (bounded by a harmonic potential to its equilibrium position) whose mechanical degrees of freedom have been treated quantum mechanically and included in the overall system dynamics. We have investigated the dynamical dressing process of the system induced by a nonequilibrium initial configuration. In particular, we have studied the time-dependent evolution which progressively leads the system to its local equilibrium dressed configuration. We have shown that the time-dependent interaction energy oscillates with time, and that at specific time intervals it is larger than the corresponding stationary value. This suggests that probing dynamical interactions could make it easier to detect corrections to the Casimir energy induced by the quantum fluctuations of the position of a macroscopic object such as a mirror. We have then investigated the long-time limit of the dynamical interaction energy, and shown that local quantities such as the field-mirror interaction energy converge to their stationary value, as obtained with a time-independent approach, while global quantities such as the field energy do not. Such a limit has been discussed for both cases of a finite and an infinite distance of the movable mirror from the fixed one. In the finite distance case, we have found revivals of the interaction energy, that represent a peculiar aspect of the time evolution of our system, and discussed their physical origin. Finally, we have generalized our results to the case of two cavities sharing the same mobile conducting wall; this has allowed us to include the effect of vacuum field fluctuations present outside the cavity. At the second order in the coupling constant, the cavity fields at the two sides of the movable wall contribute independently to the dynamical dressing process of the wall. We have argued that this independence is lost at the fourth order in the coupling.

\section*{ACKNOWLEDGMENTS}
The authors wish to thank Pablo Barcellona for stimulating and interesting discussions on the subject of this paper. F. A. acknowledges the Marie Curie Actions of the EU's 7${\mbox{th}}$ Framework Programme under REA (Grant No. 317232) for the financial support. S. B. acknowledges support from the EPSRC CM-CDT Grant No. EP/G03673X/1. M. S. K. was supported by a Leverhulme Trust research grant Project No. RPG-2014-055.

\end{document}